\documentclass{acm_proc_article-sp}
\makeatletter
\let\@copyrightspace\relax
\makeatother

\usepackage{courier}
\usepackage{color}
\usepackage{mathrsfs}
\usepackage[bf]{caption}
\DeclareCaptionType{copyrightbox}
\usepackage{amsmath}
\usepackage{amsfonts}
\usepackage{graphicx}
\usepackage{url}
\usepackage{subfig}
\usepackage{multirow}
\usepackage{tabulary}
\usepackage{colortbl}

\begin{document}

\title{Analyzing User Activities, Demographics, Social Network Structure and User-Generated Content on Instagram}

\numberofauthors{3} 
%
\author{Lydia Manikonda \qquad Yuheng Hu \qquad Subbarao Kambhampati \\ Department of Computer Science, Arizona State University, Tempe AZ 85281\\
\{lmanikon, yuhenghu, rao\}@asu.edu
}

\maketitle
\begin{abstract}
Instagram is a relatively new form of communication where users can instantly share their current status by taking pictures and tweaking them using filters. It has seen a rapid growth in the number of users as well as uploads since it was launched in October 2010. Inspite of the fact that it is the most popular photo sharing application, it has attracted relatively less attention from the web and social media research community. In this paper, we present a large-scale quantitative analysis on millions of users and pictures we crawled over 1 month from Instagram. Our analysis reveals several insights on Instagram which were never studied before: 1) its social network properties are quite different from other popular social media like Twitter and Flickr, 2) people typically post once a week, and 3) people like to share their locations with friends. To the best of our knowledge, this is the first in-depth analysis of user activities, demographics, social network structure and user-generated content on Instagram.
\end{abstract}

\section{Introduction}

Instagram, an online photo sharing, video sharing and social network service, has quickly emerged as a new medium in spotlight in the recent years. It provides users an instantaneous way to share their life moments with friends through a series of (filter manipulated) pictures and videos. Since its launch in October 2010, it has attracted more than 150 million active users, with an average of 55 million pictures uploaded by users per day, and more than 16 billion pictures shared so far~\cite{Instagram}. The extraordinary success of Instagram corroborates the recent Pew report which states that pictures and videos have become the key social currencies online~\cite{rainie2012photos}.

Despite its popularity, to date, little research has been focused on Instagram
Fundamental and critical questions such as \emph{What is instagram -- is it a social network or just an image hosting website?} \emph{What are the fundamental differences between Instagram and Twitter apart from the obvious functionalites?} \emph{How popular is Instagram?} and \emph{What exactly do users do on Instagram?} remain open and untouched. We advocate that Instagram deserves attention from web and social media research community that is comparable to the attention given to Twitter and other social media platforms~\cite{naaman2010really, ellison2007benefits}. 

\begin{figure}[ht]\vspace{-2mm}
\centering
\includegraphics[scale=0.20]{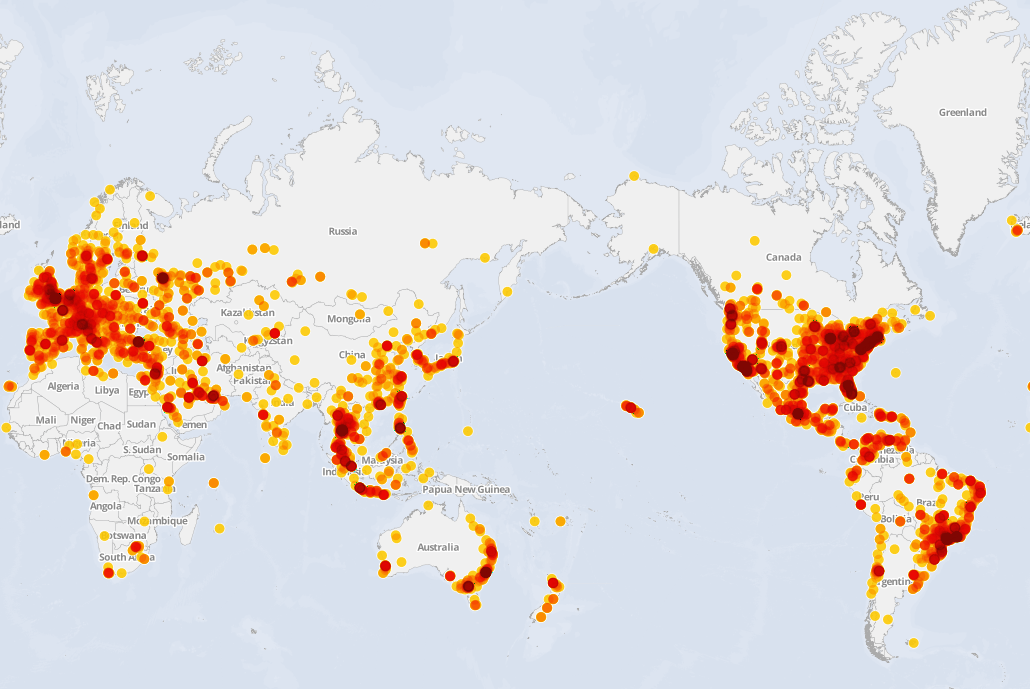}\vspace{-3mm}
\caption{Geo-locations of Instagram users}
\label{fig:instauser}
\end{figure}\vspace{-1mm}

To address the gap, in this exploratory study, we aim to acquire an initial understanding about the network structure, user activities, demographics and their shared content on Instagram, based on a large collection of user profiles crawled using Instagram API. Based on our analysis, several insights about the network properties of Instagram, user activities, demographics and the content posted by users are revealed. Using the geo-locations from our dataset, we found that Instagram is a popular social media platform with users spanning all across the world as shown in Figure.~\ref{fig:instauser}. We find that Instagram users comprise a social network but the social network properties (such as homophily, clustering coefficient, reciprocity, etc) are quite different from other popular social media platforms like Twitter and Flickr. In terms of the popularity of Instagram, we identify that the average posting time between two pictures by a typical user on Instagram is 6.5 days. We also find that for those pictures which receive comments (by other users), they have an average of 2.55 comments per post and the comments are very short (avg. \#words per comment is 4.7). Last, we reveal that users on Instagram share their geo-locations at an order of 31 times much higher when compared to Twitter.

To the best of our knowledge, we believe this is the first paper to conduct an extensive and deep analysis of Instagram's social network, user activities, demographics, and the content posted by users on Instagram. The rest of this paper is organized as follows: we begin with a general introduction of Instagram. We then present basic user statistics. Later, we present our study on the social network properties of Instagram followed by an investigation on the user generated content and the geographical aspects of Instagram. We conclude in Section 7.


\section{Background and Dataset}
\label{sec:background}
\begin{figure}[htbg]
  \centering\vspace{-2mm} \hspace{-7mm} \subfloat[]{\includegraphics[scale=0.32,
    keepaspectratio]{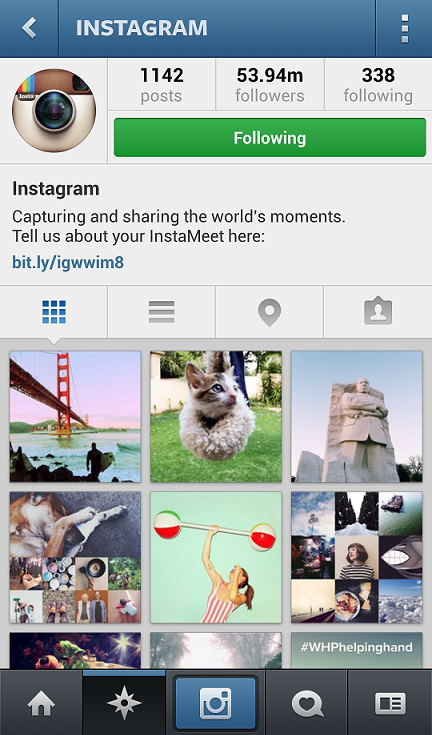}}\hspace{6mm}
  \subfloat[]{\includegraphics[scale=0.306,
    keepaspectratio]{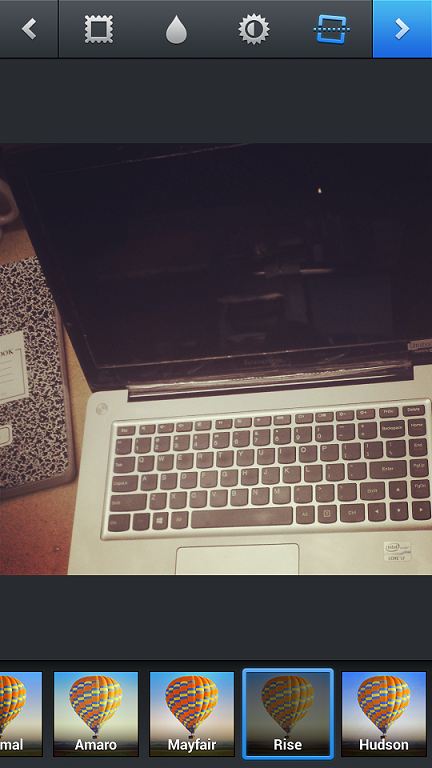}} \hspace{-9mm} \vspace{-2mm}
\caption{Interfaces of Instagram. (a) Instagram app homepage, (b) Transforming a picture using filters}
\label{fig:insta}\vspace{-3mm}
\end{figure}

Instagram (Fig~\ref{fig:insta}) is a popular photo-sharing, video-sharing social media service, with more than 150 million registered users since its launch in October 2010. It offers its users a unique way to post pictures and videos using their smartphones, apply different manipulation tools -- 16 filters -- in order to transform the appearance of an image, and share them instantly on multiple services in addition to the user's Instagram page (e.g., Flickr, Facebook, Twitter, and Foursqure). It also allows users to add captions, hashtags using the \# symbol to describe the pictures and videos, and tag or mention other users by using the @ symbol (which effectively creates a link from their posts to the referenced user's account) before posting them.

In addition to its photo taking and manipulation functions, Instagram also provides similar social connectivity as Twitter that allows a user to follow any number of other users, called ``followings''. On the other hand, the users following a Instagram user are called ``followers". Instagram's social network is asymmetric, meaning that if a user \emph{A} follows \emph{B}, \emph{B} need not follow \emph{A} back. Besides, users can set their privacy preferences such that their posted images and videos are available only to the user's followers which required approval from the user to be his/her follower. By default, their images and videos are public which means they are visible to anyone using Instagram apps or Instagram website. Users consume pictures and videos mostly by viewing a core page showing a ``stream" of the latest pictures and videos from all their friends, listed in reverse chronological order. They can also favorite or comment about these posts. Such actions will appear in referenced user's ``Updates" page so that users can keep track of ``likes" and comments about their posts. Given these functions, we regard Instagram as a kind of \emph{social awareness stream} \cite{naaman2010really} like other social media platforms such as Facebook and Twitter.

\subsection{Instagram Dataset}
In this paper, to obtain a random sample of Instagram users and retrieve their public pictures, we took a multi-step approach. First, we retrieved the unique IDs of users who had pictures that appeared on Instagram's public timeline by using Instagam API, which displays a subset of Instagram media that was most popular at the moment. This process resulted in a sample of unique users. However, after careful examination of each user in this sample, we found that these users were mostly celebrities (which explains why their posts were so popular). To avoid the sampling bias, for each user in this sample, we crawled the IDs of both their followers and friends, and later merged two lists to form one unified seed user list which contained 1 million unique users. Later, we crawled instagram to obtain the public profiles of these 1 million unique users. Among these users, only 369,828 shared their data publicly. Therefore, we retrieve complete information about these users, including their biographies, recent 20 pictures and their filters and comments, users' followers and friends (followings) lists, and users' geo-locations, if any. Our final data contains 5,659,795 pictures, 2,337,495 pictures' comments, 1,064,041 post's GPS coordinates. Note that some users may have less than 20 posts in their total activity till date.  

\section{Basic User Statistics}
\label{sec:usrstats}
\subsection{Followers and Following}
\begin{figure}[ht]\vspace{-2mm}
\centering
\includegraphics[scale=0.35]{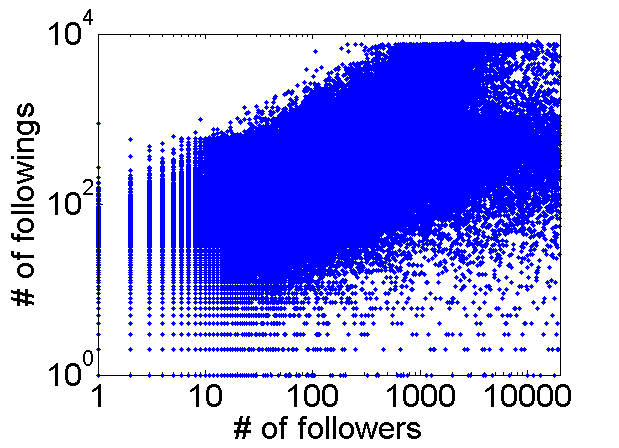}
\caption{log-log plot for followings and followers}\vspace{-3mm}
\label{fig:foll}
\end{figure}

We analyze the Instagram user network based on the followers and following lists. Figure~\ref{fig:foll} shows the log-log plot of correlation between the followers and followings. Each point in this graph represents a unique user with \emph{x}-axis representing the user's followers and \emph{y}-axis representing the user's followings. The plot clearly shows that users with a medium number of followers typically has almost the same number of followings and as the followers kept increasing, followings are also linearly increasing. At both the ends there is a high variance indicating the presence of the extreme types of users like people who follow high number of people and has less followings; people like celebrities (bottom-right) who follow very few other people but have high followers.

\subsection{Media}
\begin{figure}[htbg]
  \centering\vspace{-2mm} \hspace{-7mm} \subfloat[]{\includegraphics[scale=0.25,
    keepaspectratio]{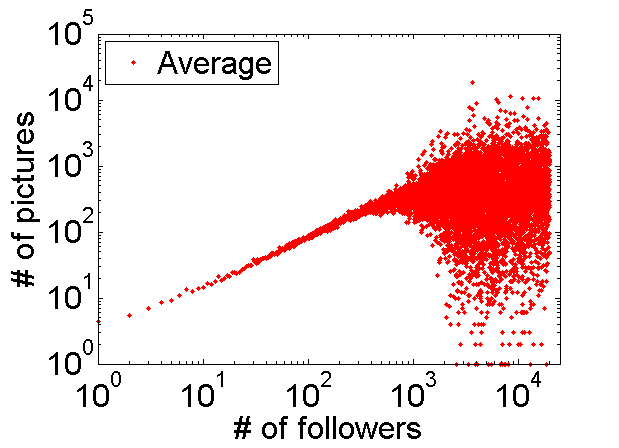}}\hspace{2mm}
  \subfloat[]{\includegraphics[scale=0.25,
    keepaspectratio]{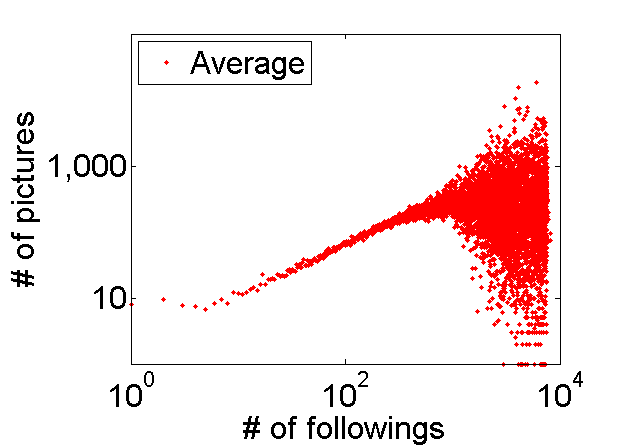}} \hspace{-9mm} \vspace{-2mm}
\caption{ Correlations on Instagram. (a) \#followers vs pictures posted, (b) \#followings vs pictures posted}
\label{fig:folfolmed}\vspace{-3mm}
\end{figure}

Next, to understand the correlation between the followers/following and the number of pictures the users post, we plot the number of pictures posted (on \emph{y}-axis) and number of followers (on \emph{x}-axis) in Figure~\ref{fig:folfolmed}(a) and we plot the number of pictures posted (on \emph{y}-axis) and number of users the users are following (on \emph{x}-axis) in Figure~\ref{fig:folfolmed}(b). To plot the graphs, we bin the followers/followings and find the average across each bin. From these log-log plots we can clearly notice that these are power-law distributions till a certain value of $x$. For users who have less than 3 followers they either post one or no picture as their average is very low. The power law is broken as the variance is very high when $x$ > 800 which is relatively low compared to the other social media like Twitter~\cite{Sue2010} which requires atleast 5000 users as followers or followings in order to see the high variance of the correlation distribution.

\subsection{Frequency of Posting Pictures}
\begin{figure}[htbg]
  \centering\vspace{-2mm} \hspace{-7mm} \subfloat[]{\includegraphics[scale=0.25,
    keepaspectratio]{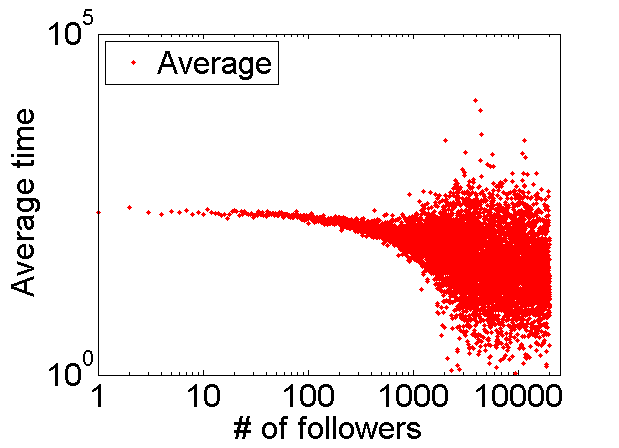}}\hspace{2mm}
  \subfloat[]{\includegraphics[scale=0.25,
    keepaspectratio]{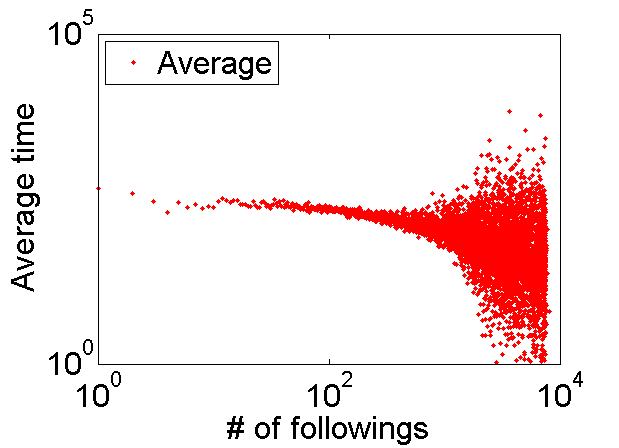}} \hspace{-9mm} \vspace{-2mm}
\caption{ Correlations on Instagram. (a) \#followers vs Frequency, (b) \#followings vs Frequency}
\label{fig:folfoltime}\vspace{-3mm}
\end{figure}

In this section we show how the frequency of posting pictures by users is correlated with the number of followings or followers. Log-log plots in Figure~\ref{fig:folfoltime}(a) and Figure~\ref{fig:folfoltime}(b) represents the correlation between the average interval between posting two pictures and followers and followings respectively. Each picture posted by users have a timestamp giving information about the time when a picture was posted. For each user we consider the sequential timestamps of the pictures posted starting from the most recent posted timestamp and then compute the difference between consecutive timestamps. For example, a user posted $k$ pictures whose timestamps are $t_1, t_2, \ldots, t_{k}$ and $t_1$ is the most recent and $t_{k}$ is the least recent. From these, we compute the relative time difference between two successive posts $\nabla$t=$t_{i}-t_{i+1}$, where $i \in [1,k-1]$. And we convert this time to hours and average across each user. It is interesting to find out that the average time between two posts by a typical user on Instagram is 6.5 days.

\subsection{Biography}
\begin{figure}[ht]
\centering
\includegraphics[scale=0.18]{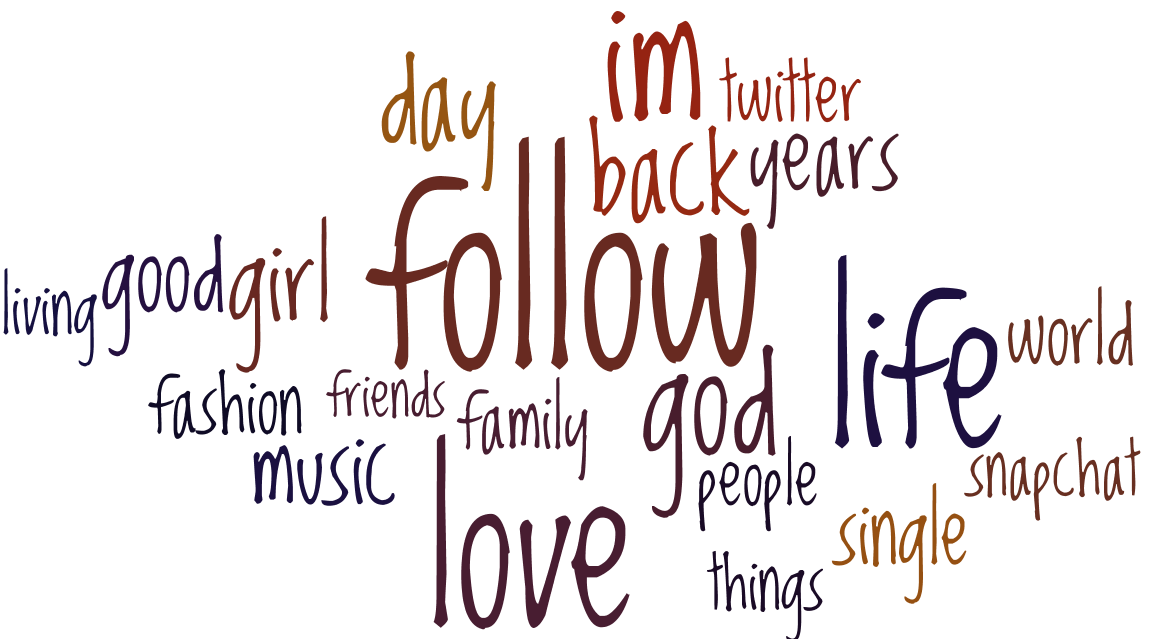}
\caption{Top unigrams extracted from the users bios}\vspace{-2mm}
\label{fig:bio}
\end{figure}

As a part of our analysis, we also focused on analyzing the biographical content of the users. Users are free to write anything as a part of the user bio. From the 369,828 users we identified the different words which the users have mentioned as part of their bio. We extracted the unigrams from this text data to identify what kinds of information do the users provide. It is interesting to notice that most of the users share their interests in the bio and it is also common to find users asking other users to follow them. Among all the unigrams, \textit{follow}, \textit{love}, \textit{life} are the top 3 unigrams with largest frequency which are 3-times the frequency of other unigrams.  In Figure~\ref{fig:bio} we show some of the top unigrams obtained from this analysis.
\vspace{-3mm}


\section{Social Network Analysis}

In this section we look at the properties of social network like homophily, reciprocity, clustering coefficient, etc to find out if the instagram network is any different from the other social media networks. We represent the network of Instagram by a directed graph \emph{G} (since it is assymetric) and denote its vertices by \emph{V} (contains $v_1, v_2, \ldots, v_n$) and edges as \emph{E} (contains $e_{11}, e_{12}, \ldots, e_{nn}$) and the edges between node \emph{i} and \emph{j} are $e_{ij}$ and $e_{ji}$ which are not the same.

\subsection{Homophily}
\begin{figure}[htbg]\vspace{-3mm}
  \centering\vspace{-2mm} \hspace{-10mm} \subfloat[]{\includegraphics[scale=0.22,
    keepaspectratio]{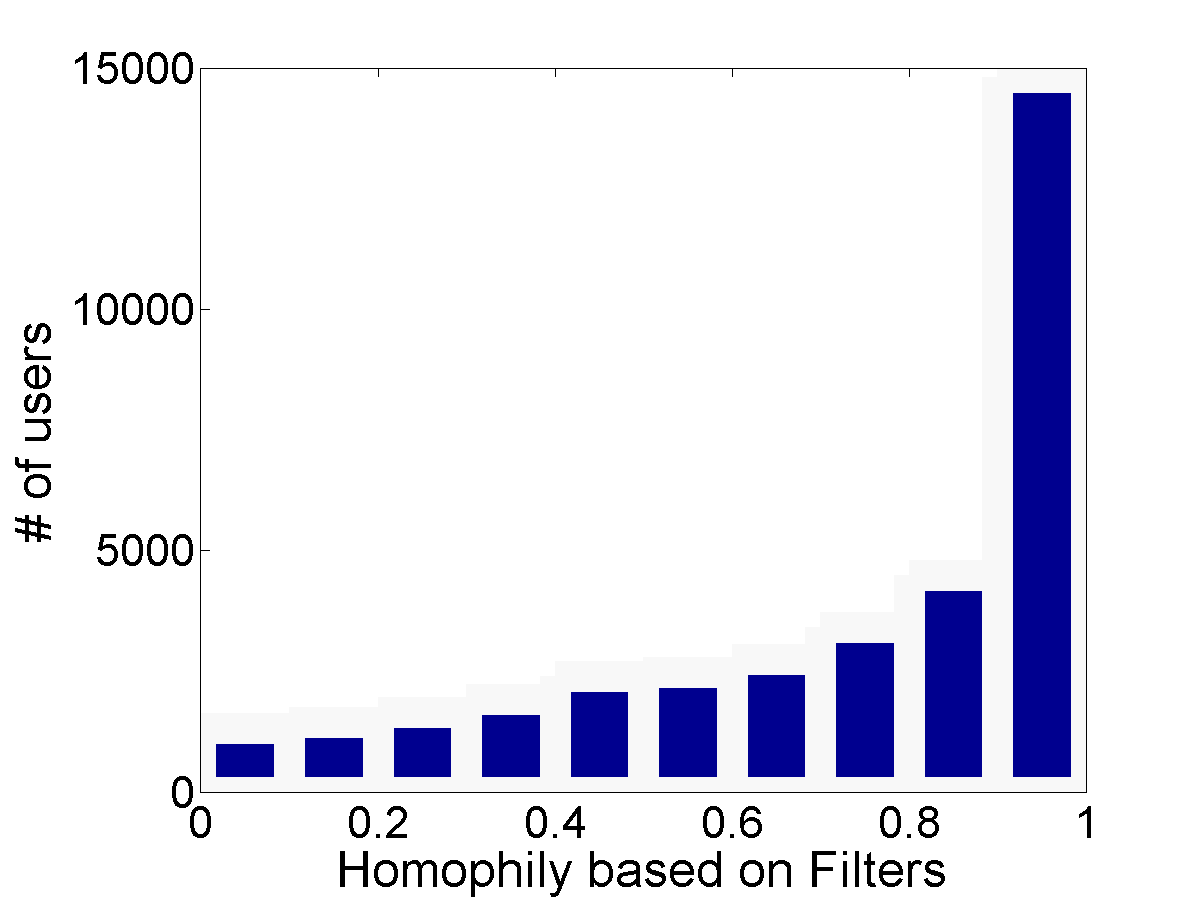}}
  \subfloat[]{\includegraphics[scale=0.22,
    keepaspectratio]{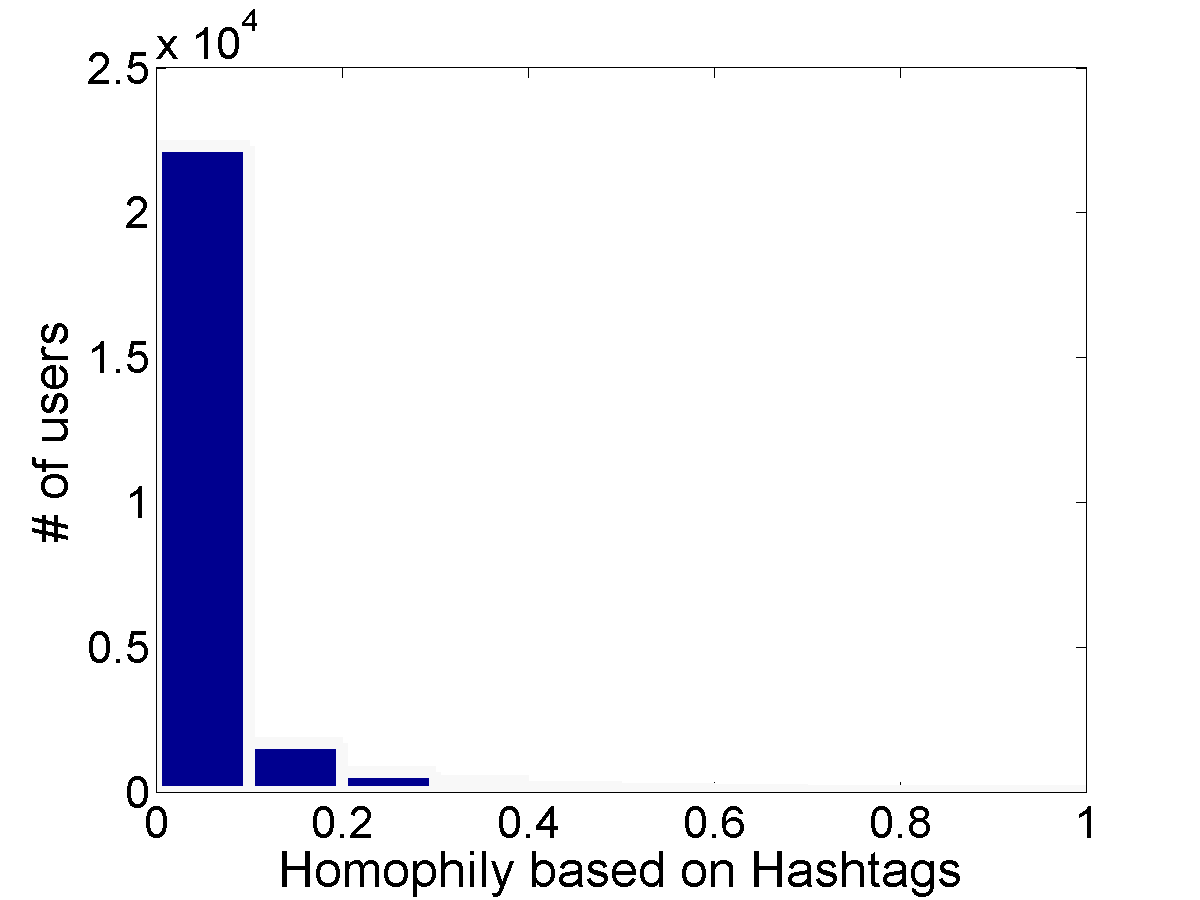}} \hspace{-13mm} \vspace{-2mm}
\caption{Homophily on Instagram. (a) Filters, (b) Hashtags}
\label{fig:homo}\vspace{-3mm}
\end{figure}

Homophily is a property which states that similar people are connected at a higher rate compared to the dissimilar people~\cite{mcpherson2001birds}. In this section we investigate homophily from two different perspectives of user's content on Instagram. We first consider the pairs of users who are reciprocally linked together in the graph $G$. For each of these user pairs we measure homophily based on the content shared in two different contexts: filters and hashtags. We extract filters used by each user and measure the similarity between the users of a pair. We use cosine distance metric to compute the similarity between reciprocated users interms of the filters used as well as hashtags. We plotted these values in Figure~\ref{fig:homo} which shows that homophily is higher in the context of using filters but completely different in the context of using hashtags. This is because users on Instagram use large number of unique hashtags where as the set of filters is fixed and cannot be changed by a user.

\subsection{Reciprocity}
\label{sec:recpro}
As similar to other social networks, celebrities are the top users in terms of the followers and most of them do not follow their followers back. We measured the reciprocity across the users of Instagram and found that it has only 14.9\% of reciprocal relationship between the users. Previous studies on other social networking services have much higher reciprocity: 68\% on Flickr~\cite{Cha2009} and 84\% on Yahoo! 360~\cite{Kumar2006} and on Twitter it is 22.1\%~\cite{Sue2010}. It is very interesting to note that instead of users explicitly asking other users to follow them inorder to be followed, the reciprocal relationship is still very low compared to other social networks.

\subsection{Clustering Coefficient}
Clustering coefficient gives a better understanding about the density of network connections in a graph. In other words, clustering coefficient measures the extent to which a user's friends are also friends of each other. This measure can be computed across all the nodes in a given graph and is averaged over all the nodes in the graph. For all the vertices $v_i$ in the graph, we compute the local clustering coefficient and finally average them to get the global clustering coefficient. For each node, clustering coefficient is measured as $C_i = \frac{|e_{jk} : v_j, v_k \in N_i, e_{jk} \in E|}{(k_{i})*(k_{i}-1)}$, where $N_i$ is the neighborhood of the $i$th node and $k_i$ is the total number of nodes present in the neighborhood $N_i$. The overall clustering coefficient is measured by averaging across all the nodes in the graph \emph{G}. Other network services like Twitter~\cite{akshay2007java} has a very low clustering coefficient of 0.106 and whereas for Instagram it is 0.42. It shows that Instagram network has more number of smaller cliques compared to Twitter. This can be compared with the very low reciprocity value obtained in the previous section which shows that there exists large number of uni-directional links in Instagram social network. 

\section{content analysis}
In this section, we describe the results obtained over analyzing the text content present in the form of picture titles, hashtags, comments, etc. As mentioned earlier, we have crawled a total of 5,659,795 pictures from the 369,828 users, where from each user we collect a maximum of only 20 images. The reason to set this threshold is to have an unbiased dataset where instead of a particular type of user accounting by 1,000 pictures and another user sharing only 25 pictures, we wanted to have a uniform collection of images from all the users. Here, we present the statistics about  filters, comments and hashtags that are associated with the pictures.


\subsection{Comments}
The main aim of this analysis on comments is to identify the average word count and the averge number of characters used by the Instagram users. Unlike twitter where for each tweet there is a maximum limit of 140 characters, Instagram doesn't have any such constraints on the character length of comments or titles. From the entire picture dataset, we extracted the comments of each picture where picture titles are not considered as comments. From the total of 5,659,795 pictures, 2,337,904 pictures don't even have a single comment which contributes to the statistics that only 41.3\% of the pictures have comments.

Given the list of all users $U$, for each user $u_i$ we compute the average number of comments ($c_i$), average number of words per comment ($w_i$) and the average number of characters per comment (${ch}_i$). If $u_i$ has a total of ${im}_i$ images that have atleast one comment, average number of comments per image for this user will be ${C_i}/{im}_i$ where $C_i$ is the total number of comments across all the images for this user. The average number of words and characters per comment for a user is computed as ${w_i}/C_i$ and  ${{ch}_i}/C_i$. Once these values are computed for all the users, we perform an average across all the $n$ users $\frac{\sum_{i=1}^{n} {C_i}/{im}_i}{n} $, $\frac{\sum_{i=1}^{n} {w_i}/C_i}{n} $ and $\frac{\sum_{i=1}^{n} {ch_i}/C_i}{n} $ to compute the average number of comments, words per comment and characters respectively. These average values for the pictures posted on instagram are shown in Table.~\ref{tab:comments}. Inspite of having no constraints on the maximum number of characters that a comment can have, users tend to post short comments.

\begin{table}[ht]\vspace{-2mm}
\centering
\begin{tabular}{|c|c|}
\hline
\textbf{Title}&\textbf{Avg. across all users}\\ \hline\hline
\#comments per picture&2.55\\ \hline
\#Words per comment&4.7\\ \hline
\#Characters per comment&32\\ \hline
\end{tabular}\vspace{-3mm}
\caption{Statistical analysis on comments}\vspace{-3mm}
\label{tab:comments}
\end{table}

\subsection{Hashtags}
In Instagram, hashtags play a key role which allows users to search for pictures that are tagged with a specific hashtag. Users provide hashtags to the pictures so that they can appear on public timeline search. This also increases the chance of identifying people with similar interests and make more connections. Frequently posted hashtags by the users can provide indirect information about the interests of users.

We have obtained all the hashtags attached with all the pictures in our database and identified the top hashtags. We observed that not all pictures in our dataset are associated with hashtags. Top-10 hashtags obtained from our analysis is shown in Table.~\ref{tab:hashtags}. It is interesting to note that users on Instagram use the top hashtags together while posting pictures to get more number of followers. Besides, users on Instagram share their sentiments and the theme of the picture through hashtags (e.g., love, beautiful, photooftheday). 

\begin{table}[ht]\vspace{-2mm}
\centering
\begin{tabular}{|c|c|}
\hline
\textbf{Hashtag} & \textbf{Frequency}\\ \hline \hline
love & 152,930\\ \hline
follow & 101,048\\ \hline
instagood & 72,658\\ \hline
me & 61,303\\ \hline
like & 59,995\\ \hline
tbt & 56,636\\ \hline
cute & 54,736\\ \hline
photooftheday & 54,444\\ \hline
beautiful & 50,602\\ \hline
happy & 49027\\ \hline
\end{tabular}\vspace{-3mm}
\caption{Top-10 hashtags of Instagram}
\label{tab:hashtags}
\end{table}\vspace{-3mm}



\subsection{Filters}
Filters on Instagram are the different lighting or color adjustments made to pictures for getting a look of low-end film cameras. Initially when launched in October 2010, Instagram forced users to use atleast one of the filters provided that can change the quality of pictures. But later they added a default filter \textit{normal} for all the pictures on Instagram which are the original pictures with no filters. There are a total of 20 filters that a user can apply to a picture. Multiple filters on the same picture at the same time is not allowed. Top Instagram filters obtained from our analysis is shown in table.~\ref{filters}. It is interesting to note that these top-5 filters are actually present in the first 7 filters of Instagram GUI. A user can see these 5 filters among the first 7 filters present in the interface of Instagram. Studying the correlations between the GUI interface design and the user's choice of filter type is beyond the scope of this paper.

\begin{table}[ht]
\centering
\begin{tabular}{|c|c|}
\hline
\textbf{Popularity} & \textbf{Filter Type}\\ \hline \hline
1&Normal \\ \hline
2&Amaro \\ \hline
3&X-Pro II \\ \hline
4&Valencia \\ \hline
5&Rise \\ \hline
\end{tabular}\vspace{-3mm}
\caption{Top-5 filters of Instagram}\vspace{-3mm}
\label{filters}
\end{table}

\section{Geolocations of Instagram}
\label{sec:geo}
Lastly, we examine the geographical aspects of Instagram. Instagram is a mobile application which is available exclusively on smart phones \footnote{Instagram does have a web interface but it only provides few functionalities compared to its mobile version.}. Given its mobile nature, we are curious about how people's location information (in terms of GPS coordinates) is embedded in their shared media on Instagram and how such information is different from the previously studied geographical aspects on Twitter. Therefore, specifically, our main focus here is to 1) explore how and to what extent people share their location on Instagram, and 2) provide an overall picture about how Instagram is being used by people in the world.

We first start with the geographical information sharing mechanism on Instagram. Before sharing each and every picture or video on Instagram, a user can choose to add graphical information by checking the ``Add to photo map" option, which by default is unchecked. The user then proceeds to use either the exact GPS coordinates as her current location or a near by point of interests (POIs) provided by Instagram (e.g., a Starbucks coffee shop). The latter way provides slightly less compromised privacy for the user. Such sharing mechanism is fundamentally different from Twitter. In Twitter, a user sets up her geographical information sharing strategy in the ``User settings". As a result, if the user opts in, all of her tweets will have GPS coordinates tagged (in contrast to the freedom Instagram provides -- location privacy can be managed individually).

Next, we examine how many pictures are tagged with GPS coordinates on Instagram. Surprisingly, we find that out of 5,659,795 pictures, more than 18.8\% contain location information. In addition, we find that at least 28.8\% of people have at least one of their pictures GPS tagged (97,871 out of 369,828). These two statistics indicate an order of 31 times much higher in terms of location sharing when compared to Twitter (only 0.6\% pictures in Twitter are tagged with GPS coordinates \cite{hu2013whoo}). This suggests that people like to use Instagram to record their daily lives and want to share that with their friends.

\begin{table}[ht]\vspace{-2mm}
\centering
\begin{tabular}{|c|c|}
\hline
\textbf{Rank}&\textbf{City}\\ \hline\hline
1&New York City, NY, USA\\\hline
2&Bangkok, Thailand  \\\hline
3&Los Angeles \\\hline
4&Sao Paulo, Brazil \\\hline
5&London, UK \\\hline
6&Moscow, Russia  \\\hline
7&Rio de Janeiro, Brazil  \\\hline
8&San Diego, CA, USA \\\hline
9&San Francisco, CA, USA  \\\hline
10&Istanbul, Turkey \\\hline
\end{tabular}\vspace{-3mm}
\caption{World's top-10 most popular geo-locations on Instagram}\vspace{-3mm}
\label{tab:geoloc}
\end{table}

In addition to these statistics, we are also interested in knowing the location from where people use Instagram. More specifically, at which city people tend to use Instagram more frequently. We then perform a processing to extract all the GPS locations and group them on a city level. This is a well-studied problem in the mobile computing research literature \cite{marmasse2000location,ashbrook2002learning}. More specifically, we first use an agglomerative clustering technique that starts by treating each individual GPS point as its own cluster. It then creates a new cluster by merging the two geographically nearest clusters into a new cluster and deleting the two constituent clusters. Merging continues until all the clusters are at least 150 Miles apart. As a result, each generated cluster represents one city metropolitan area. We then use Foursqure API to find the city information (i.e., names of city and country) based on the median of each cluster. The results are shown in Table~\ref{tab:geoloc} for the top-10 most popular cities on Instagram.

\section{Conclusion}
Considering the popularity of Instagram across the globe, in this paper we analyzed the content, geographical features and social network properties of Instagram. From our analysis we identified that Instagram is very different from other social media networks. Even though Flickr is also a similar photo-sharing social media application, Flickr has a high reciprocity compared to Instagram where users share high-quality pictures compared to everyday activity pictures captured by smart phones respectively. We also found that clustering coefficient is high for Instagram compared to Twitter. Also, Instagram users share their geo-locations at a much higher rate compared to Twitter users. Even though there are no constraints on the number of characters, users on Instagram post very short comments. Based on our analysis it is our insight that Instagram is an asymmetric social awareness platform.

\bibliographystyle{abbrv}
\begin{scriptsize}
\bibliography{sigproc}
\end{scriptsize}

\end{document}